\documentclass[fleqn,usenatbib]{mnras}

\usepackage{newtxtext,newtxmath}

\usepackage[T1]{fontenc}
\usepackage{ae,aecompl}

\usepackage{graphicx}	
\usepackage{amsmath}	
\usepackage{amssymb}	
\usepackage{longtable}
\usepackage{comment}

\title[ANNs for pulsar candidate selection]{Artificial neural networks for selection of pulsar candidates from the radio continuum surveys}

\author[N. Yonemaru et al.]{
Naoyuki Yonemaru,$^{1, 2}$\thanks{E-mail: 178d9005@st.kumamoto-u.ac.jp}
Keitaro Takahashi,$^{1, 3}$
Hiroki Kumamoto,$^{1, 2}$
Shi Dai,$^{2}$ \newauthor
Shintaro Yoshiura$^{1,4}$
and Shinsuke Ideguchi$^{5}$
\\
$^{1}$Kumamoto University, Graduate School of Science and Technology, Kumamoto, 860-8555, Japan\\
$^{2}$CSIRO Astronomy and Space Science, PO Box 76, Epping NSW 1710, Australia\\
$^{3}$International Research Organization for Advanced Science and Technology, Kumamoto University, Kumamoto, 860-8555, Japan\\
$^{4}$The University of Melbourne, School of Physics, Parkville, VIC 3010, Australia\\
$^{5}$Department of Astrophysics/IMAPP, Radboud University Nijmegen, PO Box 9010, NL-6500 GL Nijmegen, the Netherlands
}

\date{Accepted XXX. Received YYY; in original form ZZZ}

\pubyear{2015}

\begin{document}
\label{firstpage}
\pagerange{\pageref{firstpage}--\pageref{lastpage}}
\maketitle

\begin{abstract}
Pulsar search with time-domain observation is very computationally expensive and data volume will be enormous with the next generation telescopes such as the Square Kilometre Array. We apply artificial neural networks (ANNs), a machine learning method, for efficient selection of pulsar candidates from radio continuum surveys, which are much cheaper than time-domain observation. With observed quantities such as radio fluxes, sky position and compactness as inputs, our ANNs output the ``score" that indicates the degree of likeliness of an object to be a pulsar. We demonstrate ANNs based on existing survey data by the TIFR GMRT Sky Survey (TGSS) and the NRAO VLA Sky Survey (NVSS) and test their performance. Precision, which is the ratio of the number of pulsars classified correctly as pulsars to that of any objects classified as pulsars, is about $96 \%$. Finally, we apply the trained ANNs to unidentified radio sources and our fiducial ANN with five inputs (the galactic longitude and latitude, the TGSS and NVSS fluxes and compactness) generates 2,436 pulsar candidates from 456,866 unidentified radio sources. These candidates need to be confirmed if they are truly pulsars by time-domain observations. More information such as polarization will narrow the candidates down further.
\end{abstract}




\section{Introduction} \label{sec:intro}
Pulsars are rapidly-rotating neutron stars with ultra-strong magnetic fields. They emit weak radio beams from their magnetic poles which can be seen as pulses with extremely stable periods. They are used as tools in a wide range of physical experiments: low-frequency gravitational wave detection by regular monitoring of time-of-arrival (ToA) of pulses known as pulsar timing array \citep{Foster,Manchester2012,Jenet,Kramer2013}, test of gravitational theory \citep{Kramer2006,Berti}, nuclear physics inside neutron stars \citep{Lattimer}, studies of the galactic interstellar medium (ISM) and magnetic fields \citep{Han,Schnizeler}, etc. Since the discovery of the pulsar in 1968 \citep{Hewish}, many pulsar surveys have been performed for a half century \citep{Manchester2001,Cordes} and currently about 2,500 pulsars were found.

However, pulsar search in the time-domain is observationally and computationally expensive since we need to resolve narrow pulses with high time resolution. For example, the Parkes Multibeam Pulsar Survey \citep{Manchester2001} is a blind pulsar survey with the observation time per pointing of 35 minutes and the 2,670 pointings in the region of $50^\circ \le l \le 260^\circ$ and $|b| \le 5^\circ$, where $l$ and $b$ are the galactic longitude and latitude, respectively. Another example is the Arecibo Pulsar Survey using the Arecibo L-band Feed Array \citep{Cordes}. With the observation time per pointing of 17.1 (32.2) hours, it performed observations of 919 (865) pointings in the Galactic anti-center (the inner Galaxy), covering an area of 15.8 (14.8) deg$^2$.

In near future, an exceedingly large number of pulsars are expected to be discovered with the Square Kilometre Array (SKA) \citep{Keane}, and accordingly data volume will be enormous \citep{Smits,Levin}. Therefore, selection of pulsar candidates from radio continuum surveys, which is much cheaper and commensal with other sciences, will be useful to reduce the number of objects to perform time-domain observations. Recently, pulsar candidate selections with the spectral index and compactness \citep{Frail,Maan} and with the variance images \citep{Dai} have been studied, and \cite{Frail} has found five new pulsars from {\it{Fermi}} Large Area Telescope unassociated sources.

In this work, we apply artificial neural networks (ANNs) to selection of pulsar candidates from radio continuum survey data. ANN is one of the machine learning methods, which is inspired by human brain structure. Recently, machine learning methods including ANNs have been studied and applied in the field of astronomy. Some representative examples include morphological classification of galaxies \citep{Storrie-Lombardi,Naim,Folkes,Banerji}, detection and parameter estimation of gravitational waves with the multiple interferometers \citep{George}, improvement in the the accuracy of photometric redshift estimation with spectroscopic and photometric data of galaxies \citep{Collister,Vanzella,Samui}, and extraction of astrophysical parameters from the power spectrum of 21cm-line from the epoch of reionization \citep{Bukuro}. 

We construct ANNs which output the ``score'' that represents the similarity of an observed object to a pulsar from several quantities obtained from radio continuum surveys such as flux, spectral index, sky position and compactness. Using the existing radio catalog, we select radio sources which are very likely to be pulsars. This could make pulsar search much more efficient than blind surveys. The ANNs are trained with known pulsars and non-pulsar objects and thus this approach is categorized as supervised machine learning. Specifically, we construct our ANNs using data from the TIFR GMRT Sky Survey (TGSS; \cite{Intema}) and the NRAO VLA Sky Survey (NVSS; \cite{Condon}), and demonstrate how precisely our ANNs select pulsar candidates. Here, we note that any kinds of pulsars can be searched by this method, irrespective of their periods and dispersion measures (DMs), because all available pulsars are used as training data.

There are several previous works on selection of pulsar candidates with ANN \citep{Eatough,Morello,Bethapudi}. These works utilize quantities from time-domain observations, such as the pulse profile, signal-to-noise ratio (SNR), width, and chi-square of fit to the theoretical DM-SNR curve. Then, ANNs are used to judge if the signal is from a pulsar or terrestrial radio frequency interference. Furthermore, several studies \citep{Zhu,Guo,Connor,Wang} adopt convolutional neural networks (CNNs) for the classification of pulsars and fast radio bursts using {\it time-vs-phase plot} and {\it frequency-vs-phase plot} as inputs. On the other hand, our ANNs are to pick up pulsar candidates from the continuum surveys without time-domain observations. Therefore, our application is in a different phase of pulsar searching from the previous works. Note that our ANNs provide not true pulsars but their candidates, and so time-domain observations are necessary for these candidates to be identified as pulsars or not.

The outline of this paper is as follows. In section \ref{sec:source}, we introduce the source catalogs which provide training data for our ANNs and unidentified objects. In section \ref{sec:ANN}, we present the architecture and training method of ANNs. In section \ref{sec:implement ANN}, we describe features used as inputs of ANNs and how to apply ANNs to selection of pulsar candidates. In section \ref{sec:result}, we test performance of trained networks, try to interpret their interiors and apply them to the unidentified objects. We give a summary and discussion in section \ref{sec:summary}.

\section{Radio Source Catalog} \label{sec:source}
In this paper, we construct ANNs using a radio source catalog developed by \cite{Gasperin}. The catalog consists of radio sources cross-matched between TGSS and NVSS described below.

{\bf{TGSS ADR1}} - The TIFR GMRT Sky Survey \citep{Intema} is a radio continuum survey at 147 MHz carried out with the Giant Metrewave Radio Telescope (GMRT). This survey covers the north sky of $\delta = -53^\circ$ visible from the GMRT (90 \% of the celestial sphere). The resolution of this survey is $25''$ and the median rms noise is 3.5 mJy beam$^{-1}$. The overall astrometric accuracy is better than $2''$ in RA and Dec, while the flux density accuracy is estimated to be $\sim$ 10 \% for most of the survey area. The higher resolution of GMRT, combined with the data reduction strategy that down-weights the short baselines, reduced both the sensitivity of TGSS to extended emission as well as the presence of artefacts along the Galactic plane due to bright, extended sources. The largest detectable angular scale in TGSS is of order a few arcmin.

{\bf{NVSS}} - The NRAO VLA Sky Survey \citep{Condon} is a radio continuum survey at 1.4 GHz carried out with the Very Large Array (VLA). This survey covers the sky north of $\delta = - 40^\circ$ (82 \% of the celestial sphere). The survey was performed with the Very Large Array (VLA) in D and DnC configurations in full polarization. However, for this work we used only Stokes I images. The resolution is $45''$ and the background rms noise is nearly uniform at 0.45 mJy beam$^{-1}$. The overall astrometric accuracy is better than $1''$ in RA and Dec. Due to the compactness of the VLA configuration used, the surface brightness of extended sources is fairly well reconstructed up to scales of $\sim 16'$. At the same time, extended and unmodeled surface brightness from the Galactic plane lower the fidelity of images at low galactic latitude.

In \cite{Gasperin}, radio sources are cross-matched and objects with a separation less than $15''$ are regarded as the same object. Besides these cross-matched sources, we use radio sources detected by only the TGSS as well. This is because pulsars with steep spectra could be dimmer than the detection limit of the NVSS and appear only in the TGSS catalog. For these sources, we allocate the upper bound on the NVSS flux and spectral index. Hereafter, we call these objects (``S'', ``M'' and ``U'' in \cite{Gasperin}) the ``Gasperin catalog'' and it has 470,052 sources.

In order to construct ANNs, we need training data set with radio sources which are already known to be pulsars or non-pulsars. To extract pulsars from the Gasperin catalog, we cross-match it with the ATNF pulsar catalog \citep{Manchester2005} and 127 sources are identified as pulsars. The ATNF pulsar catalog consists of 2,253 normal pulsars and 360 millisencond pulsars, whereas our training data includes 107 normal ones and 20 millisencond ones. Although the ratio of pulsars in the training data to the whole pulsars in the ATNF pulsar catalog is just 4.9\%, there is not significant bias in the ratio of millisecond pulsars to normal pulsars in the training data. The Gasperin catalog is further cross-matched with the Million Quasar (MILLIQUAS) catalog \citep{Flesch} which consists of various types of radio point sources such as AGN, quasars, BL Lac objects and Seyfert galaxies (radio galaxies) which are mainly observed by the SDSS \citep{Abolfathi}. As a result, 13,166 sources are cross-matched and then identified as non-pulsars. 

\begin{figure}
\centering
\includegraphics[width=\linewidth]{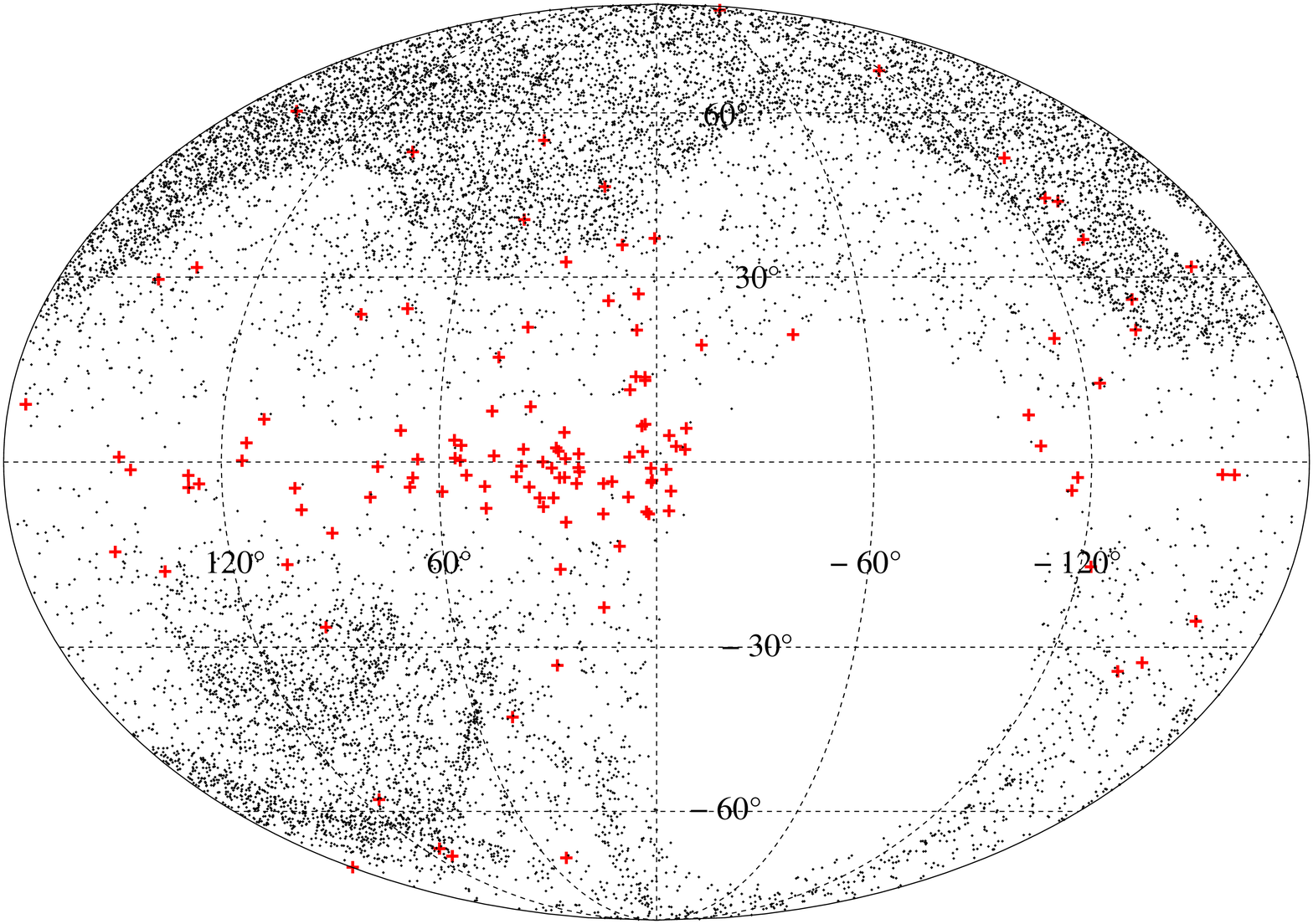}
\caption{Distribution of pulsars and non-pulsars in the galactic coordinate. The red and black crosses represent the position of pulsars and non-pulsars, respectively. Objects observed by only the TGSS are given the upper limit of 2.5mJy as the NVSS flux.}
\label{fig:psr_qso_distribution}
\end{figure}

\begin{figure}
\centering
\includegraphics[width=60mm, angle=-90]{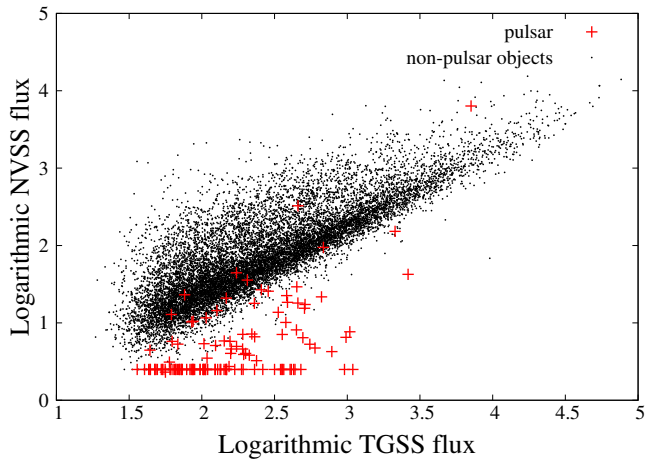}
\caption{Scatter plot of the TGSS and NVSS fluxes. The red crosses and black dots correspond to pulsars and non-pulsar objects, respectively.}
\label{fig:scatter}
\end{figure}

\begin{figure}
\centering
\includegraphics[width=60mm, angle=-90]{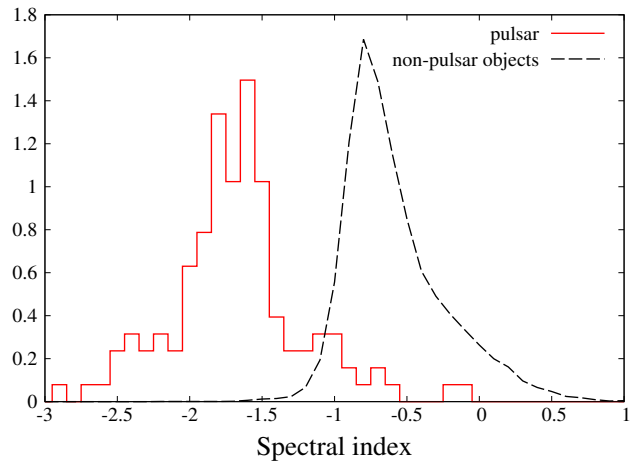}
\caption{Histogram of spectral index calculated from the TGSS and NVSS fluxes. The red solid and black dashed lines correspond to pulsars and non-pulsar objects, respectively. This histogram includes upper limits for objects observed by only the TGSS.}
\label{fig:index}
\end{figure}

\begin{figure}
\centering
\includegraphics[width=60mm, angle=-90]{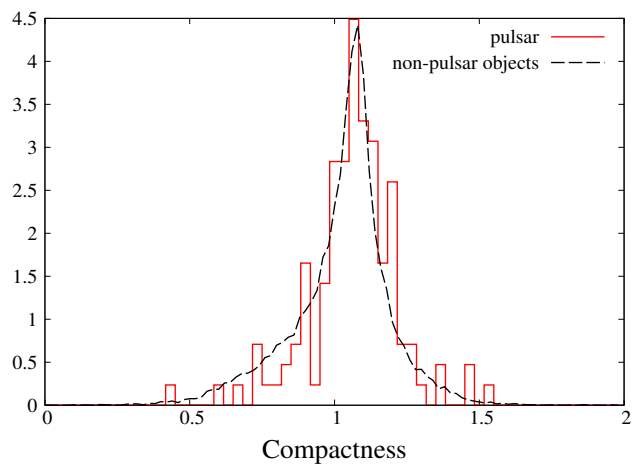}
\caption{Histograms of the compactness. The red solid and black dashed lines correspond to pulsars and non-pulsar objects, respectively.}
\label{fig:compact}
\end{figure}

Fig. \ref{fig:psr_qso_distribution} shows the distribution of the identified pulsars and non-pulsars in the galactic coordinate. The distribution of pulsars and non-pulsars are highly biased reflecting the survey region of the TGSS, NVSS and MILLIQUAS. Despite the bias of the survey region, we use these data sets as they are since unbiased data are currently unavailable. Fig. \ref{fig:scatter} shows the scatter plot of the TGSS and NVSS fluxes. We can see that pulsars have smaller NVSS flux than TGSS flux compared with non-pulsars and many of pulsars are not observed by the NVSS. This means that pulsars have steeper spectra, which can be confirmed in Fig. \ref{fig:index} which represents the histogram of spectral index. In these figures, pulsars and non-pulsars are clearly separated and these quantities will be useful to select pulsar candidates \citep{Maan}. Fig. \ref{fig:compact} shows the histogram of compactness. Although the distribution of pulsars and non-pulsars looks very similar to each other, they can still give useful information when combined with other quantities.

\section{Artificial Neural Networks} \label{sec:ANN}

ANN, which is one of machine learning methods, is a mathematical model inspired by human brain and has recently been attracting much attention. The purpose of ANN is to classify objects from input data and we need to construct a suitable network by optimizing the network parameters with training data set. In our case, the input data are observed quantities which characterize a radio source such as flux, spectral index, sky position and compactness, while the output of training data is unity/zero for a pulsar/non-pulsar, respectively. In this work, we employ the simplest model of the multilayer perceptron with three layers because of the relatively small number of input quantities mentioned above.

In this section, we describe the network architecture and the process of optimizing the network parameters (the training process) briefly.

\subsection{ANN Architectures} \label{sec:ANN_arch}

We consider ANNs which consist of three layers: the input, hidden and output layers. Each layer has neurons which are described as $x_i$, $y_j$ and $z_k$, respectively. Here, a neuron is the basic element of an ANN which generates one output from multiple inputs. An output from a neuron in the hidden layer, $y_j$, is written as
\begin{equation}
y_j = f(u_j), \label{eq:hidden}
\end{equation}
where $u_j$ is given by a linear combination of the input $x_i$, weight $w^{(1)}_{ij}$ and the bias $b^{(1)}_j$ as,
\begin{equation}
u_j = \sum_i x_i \cdot w^{(1)}_{ij} + b^{(1)}_j. \label{eq:sum of inputs}
\end{equation}
Here, $f(x)$ is the activation function and we adopt the sigmoid function, which is used commonly in the field of ANNs, given by
\begin{equation}
f(u_j) = \frac{1}{1 + \exp(-u_j)}.
\end{equation}

Concerning the output layer, an output $z_k$ is written as 
\begin{equation}
z_k = g(v_k),
\end{equation}
where $v_k$ is given by a linear combination of the output from the hidden layer $y_j$, weight $w^{(2)}_{jk}$ and the bias $b^{(2)}_k$ as,
\begin{equation}
v_k = \sum_j y_j\cdot w^{(2)}_{jk} + b^{(2)}_k \label{eq:pre_output}.
\end{equation}
In this paper, we adopt the softmax function as the activation function in the output layer,
\begin{equation}
g(v_k) = \frac{\exp(v_k)}{\sum_m \exp(v_m)}, \label{eq:softmax}
\end{equation}
which is commonly used for classification problems. In our case, the values of $z_k$ for $k=1$ and $2$ represent the scores which represent the similarity of the source to a pulsar and non-pulsar, respectively.

Here we note that, although our network includes only one hidden layer, any functional form could be approximated as long as non-linear functions are used as the activation functions and the hidden layer consists of a sufficient number of neurons. This fact is known as the universal approximation theorem \citep{Cybenko,Hornik}.

\subsection{Training}

Appropriate values of the network parameters (the weights and biases) are searched by minimizing the loss function (or the cost function) and this process is called ``training''. The loss function characterizes difference between $z_k$ obtained from the network and the correct value $t_k$. In the classification problem, the cross entropy error is often used and defined as
\begin{equation}
CE = -\frac{1}{N}\sum^N_n \sum_k t_{n, k}\log z_{n,k}, \label{ce}
\end{equation}
where $n = 1, \cdots, N$ is the number of training data. In the process of training, we need to avoid ``overfitting'', where a network is too closely fitted to the training data. There are several methods to suppress overfitting, and we adopt the weight decay method for our ANNs. The weight decay imposes a penalty on the weights and the loss function is given by the sum of the cross entropy error and the squared weights,
\begin{equation}
L = -\frac{1}{N}\sum^N_n \sum_k t_{n, k}\log z_{n,k} + \frac{1}{2}\lambda \left[ \sum_{i,j} \left(w^{(1)}_{ij}\right)^2 + \sum_{j,k} \left(w^{(2)}_{jk}\right)^2 \right], \label{loss}
\end{equation}
where $\lambda$ is a hyper parameter called the ``weight decay term'' and represents the amount of the penalty. This parameter is determined by cross validation explained in section \ref{sec:cv}.

The network parameters are optimized by the ``Momentum'' method described below. First, let $\xi (t) = (w^{(1)}_{ij}(t), b^{(1)}_j(t), w^{(2)}_{jk}(t), b^{(2)}_k(t))$ be the network parameters at $t$-th step of training. In the next step $t + 1$, they are updated as
\begin{eqnarray}
v(t+1) &=& \mu v(t) -\left. \eta \frac{\partial L}{\partial \xi} \right|_t, \label{velocity} \\
\xi (t+1) &=& \xi (t) + v(t+1), \label{update}
\end{eqnarray}
where, $\eta$ and $\mu$ are the learning rate and friction coefficient, which are fixed to 0.01 and 0.9, respectively. These are also hyper parameters and can be determined in the same way as $\lambda$. However, they affect only the efficiency of the training and not the performance of the network. Further, the number of training steps is also a hyper parameter and too many steps tend to induce overfitting. Thus, in addition to $\lambda$, we optimize the number of training steps by the method described in section \ref{sec:cv} fixing $\eta$ and $\mu$. Here, the initial values of $v(t)$ and $\xi(t)$ are set to $v(0) = 0$ and random values with normal distribution of zero mean and standard deviation of 0.1, respectively. We evaluate the derivative of the loss function in Eq.(\ref{velocity}) by the backpropagation algorithm \citep{Rumelhart}, which is a very computationally efficient method.

The training process is summarized as follows:
\begin{enumerate}
\item Initialize the network parameters $(w^{(1)}_{ij}, b^{(1)}_j, w^{(2)}_{jk}, b^{(2)}_k)$.
\item Compute output $z_k$ with Eqs.(\ref{eq:hidden}) - (\ref{eq:softmax}), and then the loss function (\ref{loss}). \label{step2}
\item Compute the derivative of the loss function with respect to the weights, and update the network parameters according to Eqs.(\ref{velocity}) and (\ref{update}).
\item Go back to \ref{step2} and iterate the number of times determined by the method explained in section \ref{sec:cv}.
\end{enumerate}

\section{Implementation of ANNs for pulsar candidate selection} \label{sec:implement ANN}

\subsection{Input parameters}

In this paper, we consider the following 7 quantities as the inputs:
\begin{enumerate}
\renewcommand{\theenumi}{\Alph{enumi}}
\renewcommand{\labelenumi}{(\theenumi)}
\item Galactic longitude $l$ normalized to [-1:1]. \label{enu:longi}
\item Galactic latitude $b$ normalized to [-1:1]. \label{enu:lati}

These quantities (\ref{enu:longi}) and (\ref{enu:lati}) represent the sky position in the galactic coordinate. The majority of pulsars are expected to be located on the Galactic plane as can be seen from Fig. \ref{fig:psr_qso_distribution} since they are formed inside the Galaxy, while extra-galactic non-pulsar objects should distribute uniformly in the sky. Thus, these quantities are potentially informative for pulsar candidate selection.

\item Absolute value of galactic latitude $|b|$ normalized to [0:1]. We consider this as an alternative to (\ref{enu:lati}), because as mentioned above pulsars are located near the Galactic plane in the sky so that $|b|$ rather than $b$ may be more useful. \label{enu:abs lati}
\item Logarithmic TGSS total flux [mJy] normalized so that the mean value is 0 and standard deviation is 0.5. \label{enu:tgss_flux}
\item Logarithmic NVSS total flux [mJy] normalized in the same way as (\ref{enu:tgss_flux}). Here, objects below the detection limit of the NVSS are given the value of upper limit of 2.5 mJy.  \label{enu:nvss_flux}
\item Spectral index $\alpha$ normalized in the same way as (\ref{enu:tgss_flux}). \label{enu:spectral_index} 

As we saw in Fig. \ref{fig:index}, pulsars tend to have steep spectra \citep{Jankowski, Ivezic, Gasperin}. Although the spectral index (\ref{enu:spectral_index}) is a direct measure of the steepness, the pair of quantities (\ref{enu:tgss_flux}) and (\ref{enu:nvss_flux}) have more information than the index and they are adopted in our fiducial network. They are specific to surveys we use in the current paper, but the fluxes at different frequencies could also be used if other radio surveys are used. Note that we assume a single power-law, but some pulsars have spectral turnover at $O(10)$ MHz and the spectra are described by a broken power-law \citep{Bilous,Murphy}. 
\item Source compactness $C$ normalized to [-1:1]. It is defined in \cite{Gasperin} as, 
\begin{equation}
C = \frac{1.071 + 2\sqrt{0.038^2 + 0.39^2\left( S_{\rm{peak}}/\sigma_l  \right)^{-1.3}}}{S_{\rm{total}}/S_{\rm{peak}}},
\end{equation}
where $S_{\rm{total}}$, $S_{\rm{peak}}$ and $\sigma_l$ are the total flux, peak flux and local rms noise of the TGSS. Pulsars with radii of a few ten km are completely point sources, while non-pulsar objects can have much more extended structures than pulsars. However, there is little difference between pulsars and non-pulsars as seen in Fig. \ref{fig:compact}. Nevertheless we consider this feature since it could have correlation with other features. \label{enu:compact}
\end{enumerate}

Then, four sets of the above quantities are taken as input parameters:
\begin{itemize}
\item[Case 1] (\ref{enu:longi}), (\ref{enu:lati}), (\ref{enu:tgss_flux}), (\ref{enu:nvss_flux}) and (\ref{enu:compact})
\item[Case 2] (\ref{enu:longi}), (\ref{enu:lati}), (\ref{enu:spectral_index}) and (\ref{enu:compact})
\item[Case 3] (\ref{enu:longi}), (\ref{enu:abs lati}), (\ref{enu:tgss_flux}), (\ref{enu:nvss_flux}) and (\ref{enu:compact})
\item[Case 4] (\ref{enu:longi}), (\ref{enu:abs lati}), (\ref{enu:spectral_index}) and (\ref{enu:compact})
\end{itemize}
where Case 1 is our fiducial set and uses original quantities, rather than derived quantities such as (\ref{enu:abs lati}) and (\ref{enu:spectral_index}). We set the number of neurons in the hidden layer as twice that of the input layer as our fiducial setup. Thus, the input, hidden and output layers have 5, 10 and 2 neurons for Cases 1 and 3, and 4, 8 and 2 neurons for Cases 2 and 4, respectively. Later, we also investigate the networks with five and fifteen neurons in the hidden layer for Case 1 (see section \ref{sec:summary}).

\subsection{Determination of Hyper-Parameters and Performance Test} \label{sec:cv}

In order to construct ANNs, we need to fix the values of hyper-parameters: the weight decay term $\lambda$ and the number of training steps. In this subsection, we describe the method to determine these hyper-parameters following \cite{Eatough}.

First, a subset is selected randomly from the whole data $(x_i, t_k)$. Here the size of the subset is typically $10\%$ of the whole data. The subset and remainder are called validation data and training data, respectively. Then, for a fixed value of $\lambda$, ANN is trained with the training data. At each step of training, the ANN is applied to the validation data and the cross entropy error is calculated between the correct value of $t_k$ and the output from the ANN. The cross entropy error tends to decrease at first but eventually turns to increase after a large number of training steps, which indicates overfitting. Thus, it is reasonable to choose the number of steps with which the cross entropy error is minimum. We repeat this process varying the value of $\lambda$ and choose both $\lambda$ and the number of steps by comparing the minima of the cross entropy error. We vary the value of $\lambda$ in the range of $-10 \leq \log_{10}\lambda \leq -2$ and consider the case of $\lambda = 0$ as well. Finally, the ANN is trained once again with all data and hyper-parameters determined in the above way. The resultant ANN is now ready to be applied to unidentified radio sources to judge if they are likely to be a pulsar or not. It should be noted that time-domain observation is necessary to confirm whether the pulsar candidates are really pulsars or not.

In this paper, we will not only apply our ANNs to unidentified sources but also demonstrate the performance of our methodology. To do the performance test as well as cross validation, we need to divide the data into three subsets: training data, validation data and test data. In our performance test, we first construct ANNs with training and validation data in the above way, and then the ANNs are applied to the test data. We repeat this process basically 10 times changing the choice of the data sets randomly. Consequently, we construct 10 independent ANNs for each Case. Note that training, validation and test data are chosen randomly every time. Finally, the performance is statistically checked and this process is commonly called the ``bootstrap'' method.

As we stated in section \ref{sec:source}, 127 pulsars and 13,166 non-pulsars were identified and they can be used as training, validation and test data. Although the ratio of the training data between pulsars and non-pulsars is imbalanced, we use 10,000 non-pulsars for training. We study cases with 200, and 1,000 non-pulsars as well to see the effect of the imbalance of the training data sets. The numbers of pulsars and non-pulsars in training, validation and test data are summarized in Table \ref{tab:num data}.

\begin{table}
	\centering
	\caption{Numbers of samples in the training, validation and test data.} 
	\label{tab:num data}
	\begin{tabular}{lcccc} 
		\hline
		 & Total & Training & Validation & Test \\
		\hline
		Pulsar & 127 & 107 & 10 & 10 \\
		Non-pulsar objects & 13,166 & 10,000 & 100 & 1,000 \\
		\hline
	\end{tabular}
\end{table}

\section{Results} \label{sec:result}

\subsection{Performance Test}
\begin{figure}
\centering
\includegraphics[width=60mm, angle=-90]{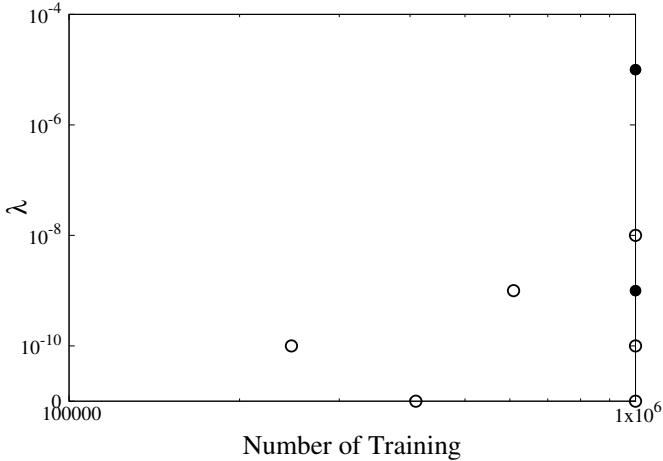}
\caption{Scatter plot of two hyper-parameters, number of training steps and the weight decay term $\lambda$, determined by the method mentioned in section \ref{sec:cv} for Case 1. The filled circles represent sets of the hyper-parameters chosen twice.}
\label{fig:cv}
\end{figure}

First, we show the results of performance tests of our ANNs. Fig. \ref{fig:cv} represents the distribution of hyper-parameters determined by the method mentioned in section \ref{sec:cv} for 10 realizations of Case 1. The number of training steps is in the range of $[10^5,10^6]$, while the weight decay term $\lambda$ is scattered in a wide range below $10^{-4}$. Hyper-parameters are determined to optimize the network to given training data and the variation of hyper-parameters is attributed to the variation of training data which are chosen randomly for each realization. Consequently, the performance of ANNs also varies among the 10 realizations. We discuss the average performance below.

\begin{figure}
\centering
\includegraphics[width=60mm, angle=-90]{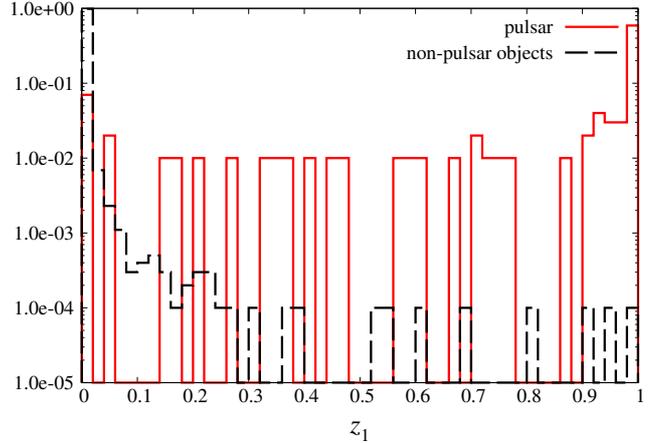}
\caption{Score distribution of $z_1$ of pulsars and non-pulsars of test data of all of the 10 realizations for Case 1.}
\label{fig:histogram}
\end{figure}

Next, we show the results of performance test of trained networks. The outputs of our ANNs are the scores, $z_1$ and $z_2$, which represent the similarity to a pulsar and non-pulsar, respectively, and the sum is normalized to unity. Fig. \ref{fig:histogram} shows the histogram of $z_1$ of the test data obtained from all of the 10 realizations for Case 1. As can be seen, the value of $z_1$ is almost zero or unity for most objects.

\begin{figure}
\centering
\includegraphics[width=60mm, angle=-90]{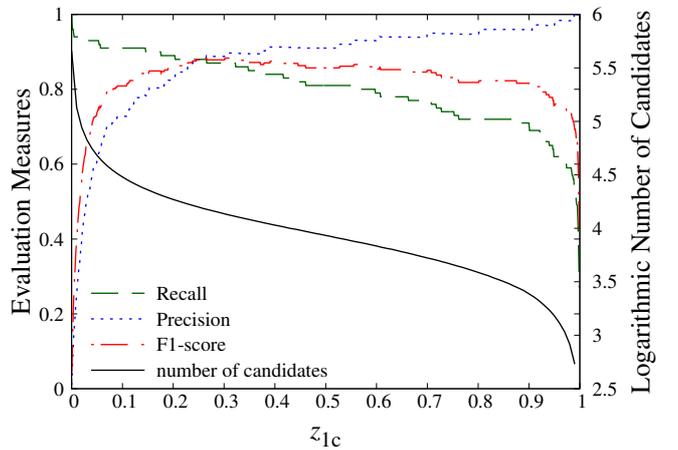}
\caption{Recall, Precision and F1-score averaged over 10 realizations (left axis) and the logarithmic number of pulsar candidates (right axis) which are obtained by applying the trained ANN to the unidentified objects in the Gasperin catalog for Case 1 against the criterion of $z_1$.}
\label{fig:evaluation}
\end{figure}

To determine the criterion of $z_1$ that classifies objects into the pulsars or non-pulsars, we use the following evaluation measures,
\begin{eqnarray}
\rm{Recall} &=& \frac{TP}{TP + FN}, \\
\rm{Precision} &=& \frac{TP}{TP + FP}, \\
\rm{F1-score} &=& \frac{2\times \rm{Recall}\times \rm{Precision}}{\rm{Recall} + \rm{Precision}},
\end{eqnarray}
where $TP$, $FN$ and $FP$ stand for true positives, false negatives and false positives which represent the numbers of pulsars classified as pulsars, pulsars classified as non-pulsars and non-pulsars classified as pulsars, respectively. Fig. \ref{fig:evaluation} shows the averaged Recall, Precision and F1-score over 10 realizations as a function of the criterion of $z_1$, $z_{1 \rm{c}}$, for Case 1. Although F1-score is the largest at around $z_{1 \rm{c}} = 0.3$, it is almost flat between 0.1 and 0.9 and drops sharply for $z_{1 \rm{c}} > 0.9$. In Fig. \ref{fig:evaluation}, the number of pulsar candidates which are obtained by applying the trained ANN to the unidentified objects in the Gasperin catalog is also shown (see section 5.4). Because non-pulsar objects are considered to dominate the unidentified objects, we choose the threshold to be $z_{1 \rm{c}} = 0.9$ which gives a high value of Precision and a relatively small number of pulsar candidates. The situation is similar for other Cases, therefore, we take $z_{1 \rm{c}} = 0.9$ as the criterion of pulsar candidates for all Cases hereafter.

Table \ref{tab:probability} shows the mean and standard deviation of the above three evaluation measures over 10 realizations for the fiducial and variant ANNs. 

Fixing the number of non-pulsar training data to 10,000, Case 3 has the largest average evaluation measures. Thus, Case 3 would be the best ANN of the four. Although, there are uncertainties in the evaluation measures over the networks, it can also be seen from the comparison of the four cases that the absolute value of galactic latitude is a better input than the galactic latitude itself, while individual fluxes of TGSS and NVSS are better than spectral index. Finally, comparing Case 1 with different numbers of non-pulsar training data, it is seen that its increase results in decreasing Recall, but increasing Precision and smaller its variance among the networks. The network in the case of the non-pulsar train data of 200 would be the best in terms of F1-score, but high Precision which means that the ratio of non-pulsars classified as incorrectly pulsars is small is also very important because the number of non-pulsar objects should be much larger than that of pulsars as mentioned above. 

Another common measure of the effectiveness of ANNs is the Area Under the ROC Curve (AUC), where the ROC curve stands for the receiver operating characteristic curve. The ROC curve is a plot of the Recall (also known as the true positive rate) against the false positive rate (FPR) at various values of $z_{1c}$. Here, the FPR is given by ${\rm{FPR}} = FP/(FP + TN)$, where $TN$ stands for true negatives. Noting that both the Recall and FPR are in a range of $[0,1]$, the AUC is defined as the area which is surrounded by the ROC curve and two lines of Recall=0 and FPR=1. The AUC can take a value from zero to unity, and a classifier with a larger value of the AUC is considered to be more powerful. The average values of the AUC are also shown in Table \ref{tab:probability}. These values are generally very close to unity and show the high performance of the trained ANNs.

\begin{table*}
	\centering
	\caption{Recall, Precision and F1-score with a pulsar-candidate criterion of $z_1 \ge 0.9$, average value of the AUC and the number of pulsar candidates obtained by applying the trained ANN to the unidentified objects in the Gasperin catalog with $z_1 \ge 0.9$.}
	\label{tab:probability}
	\begin{tabular}{lcccccc} 
		\hline
		 Input & \multicolumn{3}{c}{Case 1} & Case 2 & Case 3 & Case 4 \\
         Number of Non-pulsar Training data & 200 & 1,000 & 10,000 & 10,000 & 10,000 & 10,000 \\
		\hline
		Recall (\%) & 85.0$\pm$14.9 & 79.5$\pm$16.1 & 71.0$\pm$17.9 & 52.0$\pm$16.9 & 75.0$\pm$14.3 & 65.0$\pm$19.0 \\
		Precision (\%) & 92.6$\pm$12.3 & 96.9$\pm$10.9 & 96.1$\pm$6.32 & 95.4$\pm$7.47 & 98.8$\pm$3.95 & 98.3$\pm$5.27 \\
		F1-score (\%) & 87.9$\pm$11.7 & 86.6$\pm$13.0 & 80.3$\pm$14.5 & 65.3$\pm$17.9 & 84.6$\pm$9.39 & 76.4$\pm$13.4 \\
		Averaged AUC & 0.974 & 0.968 & 0.976 & 0.967 & 0.973 & 0.989 \\
		Number of candidates & 20,971 & 52,615 & 2,436 & 3,765 & 11,675 & 3,109 \\
		\hline
	\end{tabular}
\end{table*}

\subsection{Interpretation of Weights} \label{sec:weight interpretation}

In this subsection, we try to interpret the behavior of the weights and understand the interior of the trained ANNs. To do this, we neglect the activation functions for simplicity. In this approximation, the output from the hidden layer, Eq.~(\ref{eq:sum of inputs}), is given by
\begin{equation}
y_j = a \sum_i x_i\cdot w^{(1)}_{ij} + b^{(1)}_j, \label{eq:linear_hid}
\end{equation}
and by instituting this into Eq.(\ref{eq:pre_output}), we obtain
\begin{equation}
v_k = a \sum_i x_i\cdot w_{ik} + b^{(2)}_k .
\end{equation}
Here, $a \sim 0.2$ is the coefficient of the linear function, and $w_{ik}$ is the product of two weight matrices,
\begin{equation}
w_{ik} = \sum_j w^{(1)}_{ij}\cdot w^{(2)}_{jk}. \label{eq:matrix}
\end{equation}
Here, we ignore the bias $b_j$ since we focus on the behavior of the weights in the networks. We can sum up the weights with respect to the hidden layer and the input layer is connected with the output layer directly by approximating the sigmoid function as the linear one. In the following, we study behavior of this $w_{ik}$.

\begin{figure*}
\begin{tabular}{cc}
\begin{minipage}{0.5\hsize}
\centering
\includegraphics[width=60mm, angle=-90]{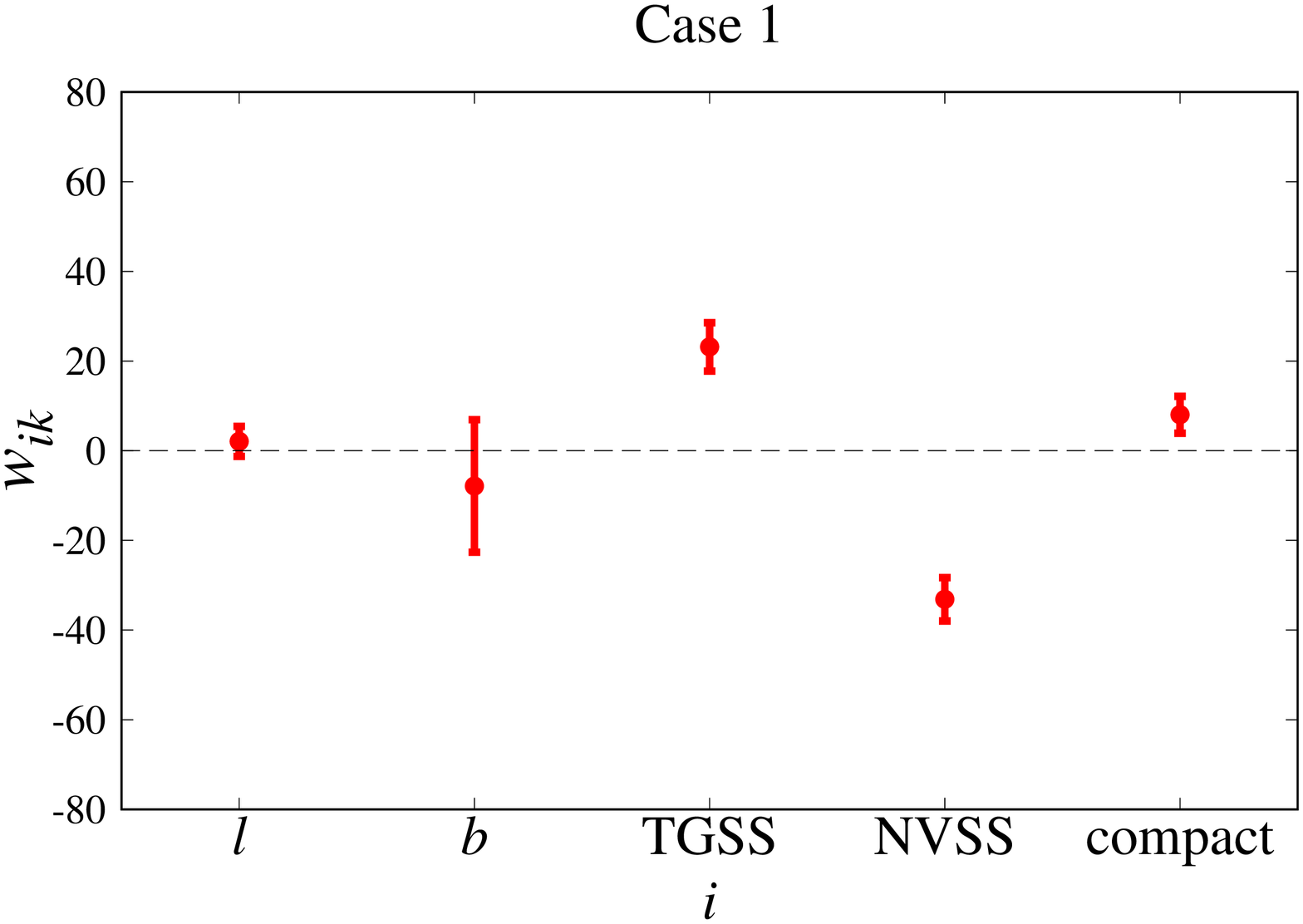}
\end{minipage}
\begin{minipage}{0.5\hsize}
\centering
\includegraphics[width=60mm, angle=-90]{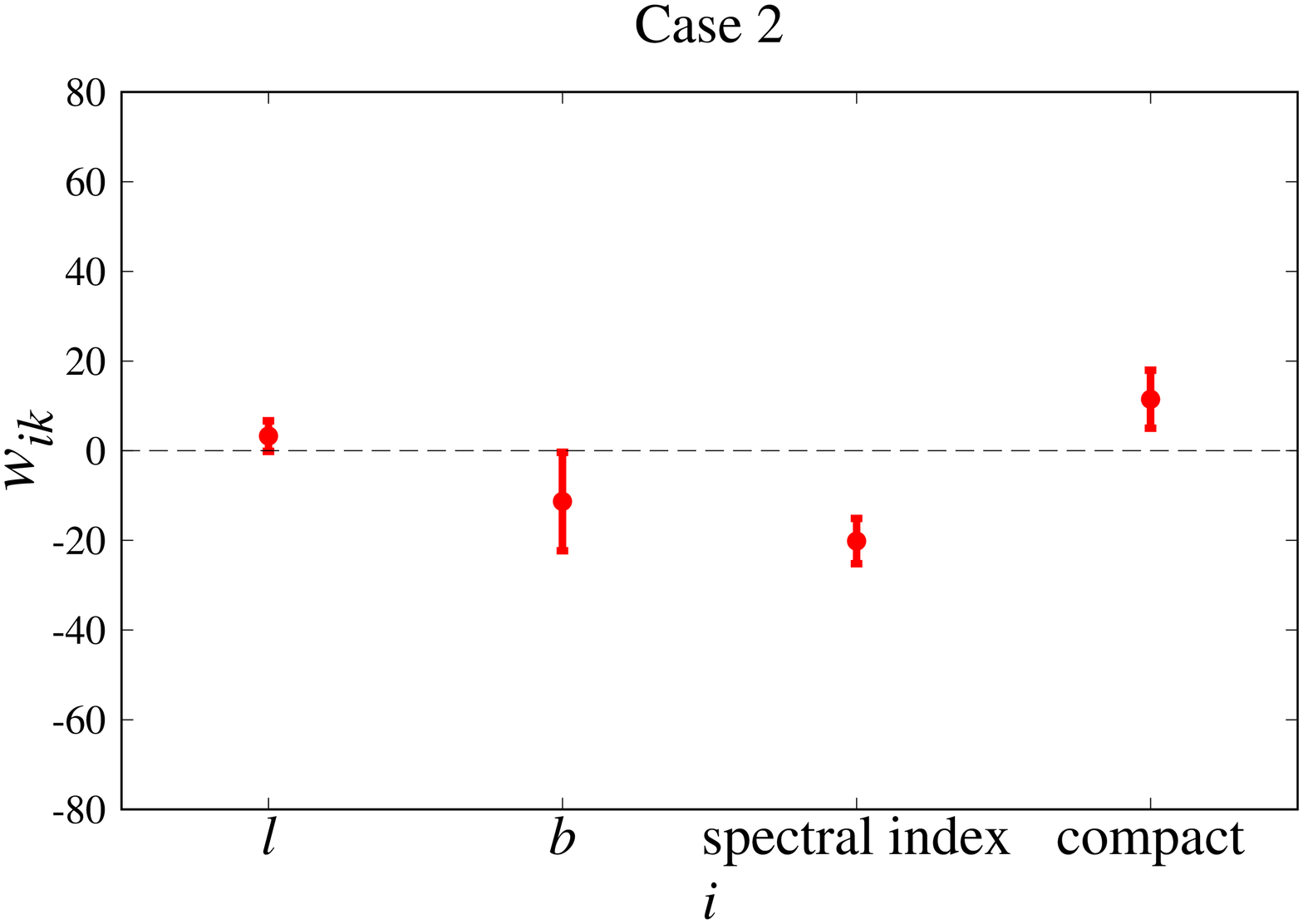}
\end{minipage}\\
\begin{minipage}{0.5\hsize}
\centering
\includegraphics[width=60mm, angle=-90]{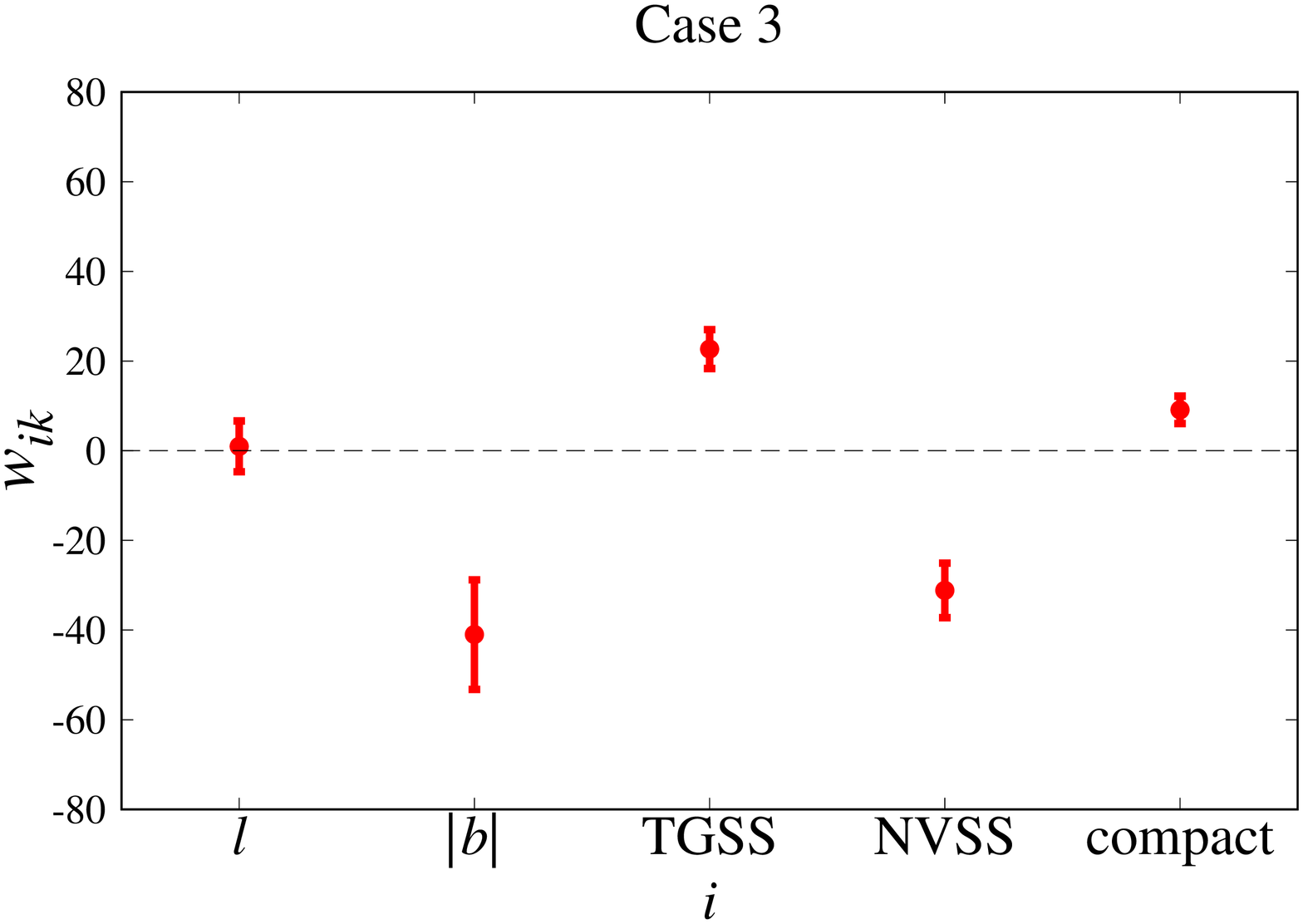}
\end{minipage}
\begin{minipage}{0.5\hsize}
\centering
\includegraphics[width=60mm, angle=-90]{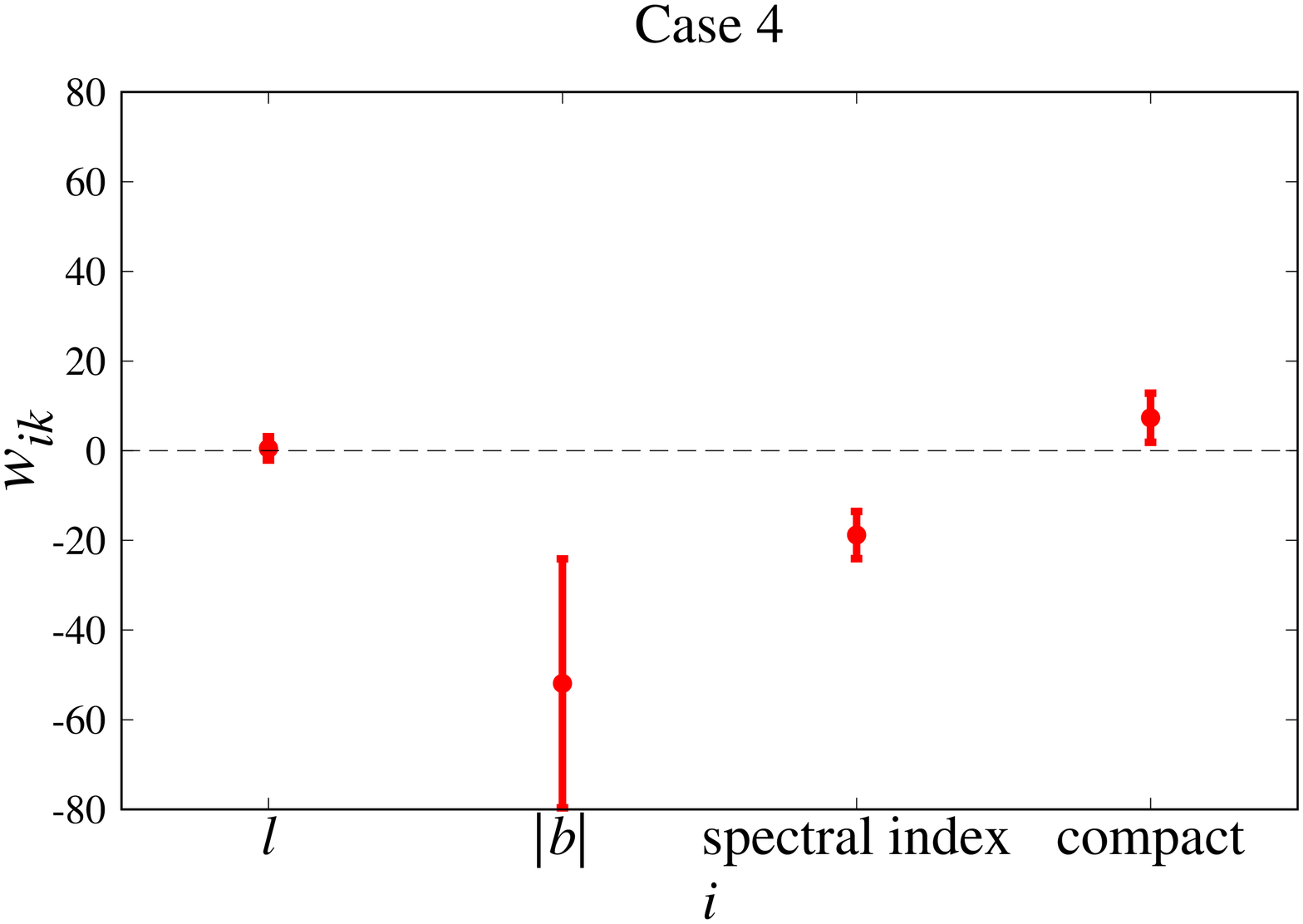}
\end{minipage}
\end{tabular}
\caption{Plot of the averaged $w_{i1}$ over realizations with the error bar for each Case.}
\label{fig:weight}
\end{figure*}

Because $w_{i2} = - w_{i1}$, we argue the behavior of $w_{i1}$ only. Fig. \ref{fig:weight} shows the mean and standard deviation of $w_{i1}$ for each Case over 10 realization. The weight of the longitude, $w_{11}$, for every Case is consistent with 0, which implies the longitude is not informative for the selection. On the other hand, the weights of the latitude are consistent with zero for Cases 1 and 2 which uses the latitude itself, while those for Cases 3 and 4, which use the absolute value of the latitude, $|b|$, are significantly negative. This implies that $|b|$ is useful to select pulsar candidates and that pulsars tend to have small value of $|b|$, that is, pulsars are mostly located within the Galactic plane.

The negative and positive weights of the TGSS and NVSS fluxes for Cases 1 and 3 indicate that an object which is bright and dark in the TGSS and NVSS, respectively, tends to be selected as a pulsar. This is consistent with the fact that pulsars have steep spectra as seen in Fig.~\ref{fig:index}. For Cases 2 and 4, this is seen as the negative weights of the spectral index.

The weight of the compactness is almost consistent with 0, but is slightly positive for all Cases. Alghough the compactnesses of the pulsar and non-pulsar objects look almost the same in Fig.~\ref{fig:compact}, this might imply the trained ANNs detect invisible correlation with other parameters.

Thus, our simple interpretation of weights is consistent with our understanding of the basic properties of pulsars and non-pulsars. However, it should be noted that, in the above interpretation, possible correlations between the input quantities are marginalized and only direct correspondence between the inputs and the pulsar score is investigated. Thus, even if the average weight is consistent with zero, it does not necessarily mean the corresponding input has no effect on the pulsar selection.

\subsection{Missed Objects}

In this subsection, we show features of the ``missed'' objects in the test process focusing on Case 1. Here, the missed pulsars represent true pulsars in the test data which have $z_1 < 0.9$ and, consequently, were not selected as pulsar candidates. Conversely, missed non-pulsars represent true non-pulsars which have $z_1 \ge 0.9$ and, consequently, are selected as pulsar candidates. As explained above, we have 10 realizations for each Case and each ANN is tested with 10 pulsars and 1,000 non-pulsars (see Table \ref{tab:num data}). Thus, the total numbers of pulsars and non-pulsars in the test data are 100 and 10,000, respectively.

Among the test data, 29 pulsars and 3 non-pulsars were missed through 10 realizations and the fraction of wrong selection is 29\% (false negative rate) and 0.03\% (false positive rate), respectively. The former fraction may look large and this is partly because the criterion of classification is rather high ($z_{1 \rm{c}} = 0.9$). However, it is important to suppress the latter as low as possible, rather than the former. This is because most of unknown objects are considered to be non-pulsars so that only a small value of the false positive rate can substantially increase the number of false pulsar candidates, which makes pulsar search in time domain very inefficient. Thus, we accept this relatively high false negative rate.

\begin{figure}
\centering
\includegraphics[width=\linewidth]{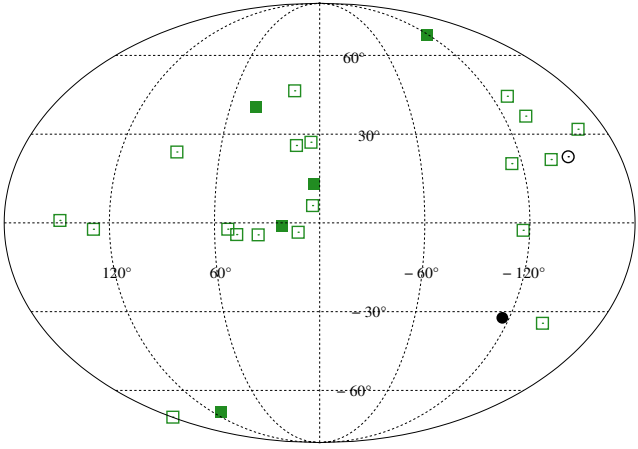}
\caption{Distribution of the ``missed'' pulsars and non-pulsar objects described by the green squares and black circles in the galactic coordinate.}
\label{fig:miss_map}
\end{figure}

Fig. \ref{fig:miss_map} shows the distribution of the missed pulsars and non-pulsars in the galactic coordinate. The filled symbols represent objects which were missed multiple times. Although most of pulsars are located in the Galactic plane as seen in Fig.~\ref{fig:psr_qso_distribution}, many of the missed pulsars distribute roughly uniformly, which indicates that pulsars at high latitudes are more likely to be missed. On the other hand, because the number of missed non-pulsars is very small, it is not possible to discuss their distribution.

\begin{figure}
\centering
\includegraphics[width=60mm, angle=-90]{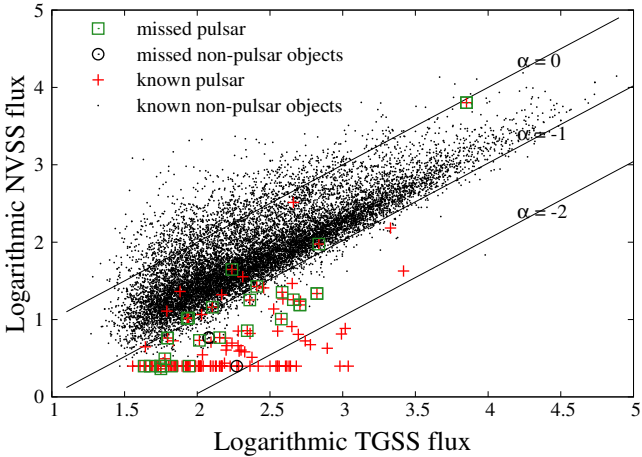}
\caption{Scatter plot of the logarithmic TGSS and NVSS fluxes of the ``missed'' pulsars and non-pulsar objects described by the green squares and black circles. Solid lines shows the spectral index contours which describe $\alpha = 0$, -1 and -2 from the top.}
\label{fig:miss_flux}
\end{figure}

Fig. \ref{fig:miss_flux} shows the scatter plot of the missed objects in a plane of logarithmic TGSS and NVSS fluxes. We see that two missed non-pulsars, one of which is missed twice, are located out of the main region of the non-pulsar population. On the other hand, while many of missed pulsars have large values (small absolute values) of spectral index, the steepest missed pulsars have spectral indeces about -1.5. Although there are more known pulsars than known non-pulsars around $\alpha \sim -1.5$, the number of known non-pulsars is not negligible ($\sim 10$). Therefore, $\alpha \sim -1.5$ would be the boundary of the classification and this is why some pulsars with $\alpha \sim -1.5$ are missed.

\subsection{Applying ANNs to unidentified objects} \label{sec:apply ANN}

We apply our trained ANNs to the unidentified objects in the Gasperin catalog. In this application, we use the networks of all Cases individually. We choose training and validation data randomly, determine the hyper parameters by the method mentioned in subsection \ref{sec:cv}, train the network with those hyper parameters, and then apply the trained network to the 456,866 unidentified objects. The number of pulsar candidates with $z_1 \ge 0.9$ for each Case is shown in Table \ref{tab:probability}. 

Comparing Case 1 ANNs which were trained with different numbers of non-pulsars (200, 1,000 and 10,000), the number of candidates is smallest for the network with the 10,000 non-pulsar training samples. On the other hand, comparing the 4 Cases with 10,000 non-pulsar training samples, Case 1 has the smallest number of candidates, while Precisions are comparable within the statistical errors. Because, as stated before, the unidentified objects would be dominated by non-pulsars, we regard the Case 1 with 10,000 non-pulsar training samples as the most effective ANN.

Comparing individual candidates of Cases 1-4, we find that, among 2,436 pulsar candidates of Case 1, 2,047 (84\%), 1,996 (82\%) and 819 (33.6\%) are common with Cases 2, 3 and 4, respectively. The similarity is relatively low for Case 4 compared with Cases 2 and 3. This would be because, for Case 4, more inputs are replaced from Case 1 compared with Cases 2 and 3.

\begin{figure}
\centering
\includegraphics[width=\linewidth]{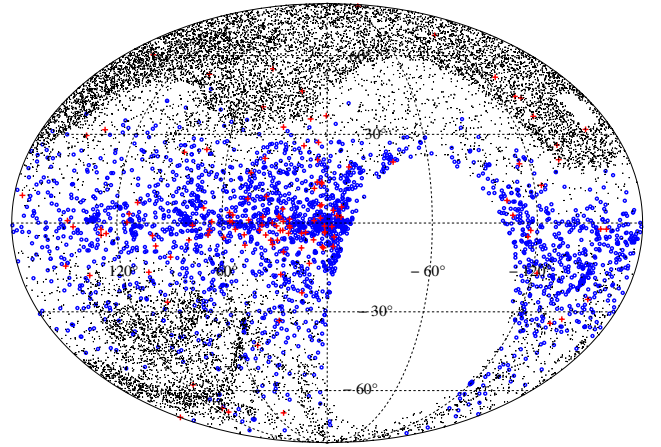}
\caption{Distribution of the known pulsars (red cross), known non-pulsars (black cross) and 2,436 Case-1 candidates (blue circle) in galactic coordinate.}
\label{fig:candidate_map}
\end{figure}

Next, we describe the candidates of Case 1 more in detail. Fig. \ref{fig:candidate_map} shows the distribution of the known pulsars, known non-pulsars and 2,436 pulsar candidates in the sky. The candidates are mainly located on the Galactic plane, but some of them are at high latitudes. This distribution seems to be biased by the SDSS-surveyed and non-observed areas especially in the upper right area ($-180^\circ \lesssim l \lesssim -120^\circ$ and $0^\circ \lesssim b \lesssim 30^\circ$) of Fig.~\ref{fig:candidate_map}, where the less candidates are distributed than in the upper left area ($120^\circ\lesssim l \lesssim 180^\circ$).

\begin{figure}
\centering
\includegraphics[width=60mm, angle=-90]{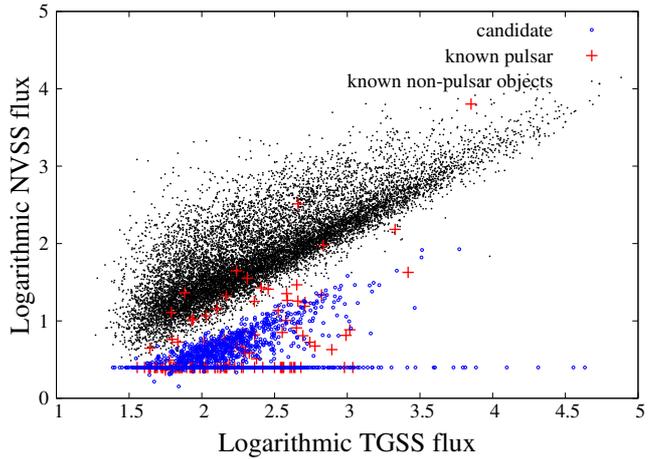}
\caption{Scatter plot of the logarithmic TGSS and NVSS fluxes of the known pulsars and non-pulsar objects, and 2,436 candidates with $z_1 \ge 0.9$ described as the red crosses, black dots and blue circles.}
\label{fig:candidate_flux}
\end{figure}

Fig. \ref{fig:candidate_flux} shows a scatter plot of the known pulsars, known non-pulsars and 2,436 candidates on a plane of the logarithmic TGSS and NVSS fluxes. The distribution of candidates mostly overlaps with that of known pulsars, which small NVSS fluxes and steep spectral indices.

To show the validity of our method, we checked if our candidates include newly-found pulsars and candidates in \cite{Maan,Frail}. We found that our candidates cross-match 21 of 25 candidates in \cite{Maan}, while 3 of 5 new pulsars and 3 of 5 candidates in \cite{Frail} are cross-matched with our candidates. In fact, these 4 non-cross-matched objects in \cite{Frail} are not included in our catalog of unidentified objects. Thus, our ANN selects all of new pulsars and candidates in \cite{Frail} included in our catalog and this fact shows the effectiveness of our method for pulsar candidate selection.

\section{Summary and Discussion} \label{sec:summary}

We applied artificial neural networks (ANNs) for efficient selection of pulsar candidates from continuum surveys. From the input quantities such as radio fluxes, sky position and compactness, ANNs were constructed to output a score that represents a degree of likelihood for an object to be a pulsar. We demonstrated ANNs based on existing survey data by the TGSS and the NVSS and tested their performance varying the input parameters and the number of training data. Finally, we obtained pulsar candidates by applying the trained ANNs to unidentified radio sources. For the validation, our candidates should be confirmed if they are truly pulsars with time-domain observation. This is ongoing and will be presented elsewhere.

We evaluated our trained networks with test data which consist of known pulsars and known non-pulsars. As a result, it is indicated that the trained networks have high classification performance and Precision, which is the ratio of the number of pulsars classified correctly as pulsars to the number of objects classified as pulsars, is basically higher than 95\%. This implies that the fraction of non-pulsars among pulsar candidates is less than 5\%, albeit non-pulsar objects are considered to dominate radio point sources. Our ANNs are also tested with pulsar candidates and newly-found pulsars in \cite{Maan,Frail}. We found that our candidates generated by Case 1 ANN include 21 of 25 candidates in \cite{Maan} and all new pulsars and candidates in \cite{Frail} contained in our unidentified catalog. Thus, our ANNs work pretty well for the pulsar candidate selection.

\begin{figure}
\centering
\includegraphics[width=\linewidth]{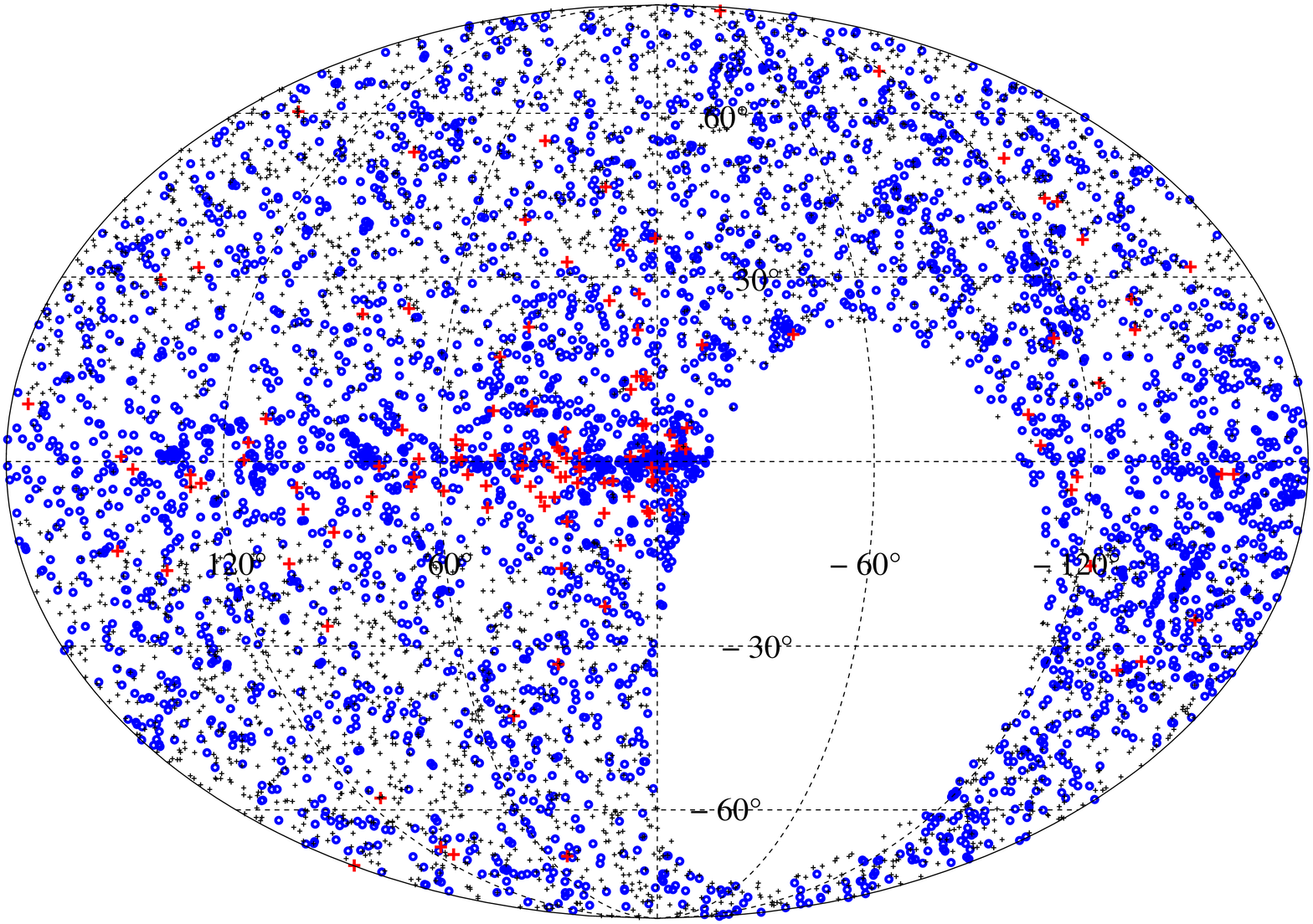}
\caption{Distribution of the known pulsars (red cross) and 3,639 non-pulsar objects sampled uniformly in the sky (black cross), and 3,636 pulsar candidates (blue circle) in the galactic coordinate.}
\label{fig:can_uniform_map}
\end{figure}

Let us discuss the effect of the spatial bias of the training data, especially non-pulsars. As we saw in Fig.~\ref{fig:psr_qso_distribution}, the non-pulsar distribution is biased due to the limited SDSS survey area. In order to investigate the effect of the spatial bias, we performed the same analysis with 3,639 non-pulsars sampled spatially uniformly in the Gasperin catalog area. We used 3,500 for training and 100 for validation out of 3,639 non-pulsars. Consequently, we obtained 3,636 candidates ($z_1 \ge 0.9$) by applying the ANN to the unidentified objects. The spatial distribution of the training data and pulsar candidates are shown in Fig. \ref{fig:can_uniform_map}. Compared to Fig. \ref{fig:candidate_map}, the distribution of the candidates extends to the outside of the Galactic plane, although the concentration on the plane could still be seen. This implies that the spatial distribution of candidates is affected by the spatial bias of the training data. Further study on this effect is beyond the scope of the current work and will be presented elsewhere. 

As mentioned in section \ref{sec:ANN_arch}, ANNs with a larger number of neurons in the hidden layer are expected to work better, while its computational cost becomes larger. Here, we briefly compare the performance of three Case-1 ANNs with 5, 10 and 15 neurons in the hidden layer. Table \ref{tab:hidden} shows Recall, Precision, F1-score, the average AUC and the number of candidates ($z_{1\rm{c}} \ge 0.9$) for the three ANNs. These characteristic numbers are almost within the statistical fluctuations, while the case with 5 neurons has slightly low performance. Hence, we conclude that it is reasonable to set the number of neurons in the hidden layer to 10 as our fiducial network.

\begin{table}
	\centering
	\caption{Same as Table \ref{tab:probability} for Case-1 ANNs with 5, 10 (fiducial) and 15 neurons in the hidden layer.}
	\label{tab:hidden}
	\begin{tabular}{lccc}
		\hline
		Number of Neurons & 5 & 10 & 15  \\
		\hline
		Recall (\%) & 56.0$\pm$19.6 & 71.0$\pm$17.9 & 67.0$\pm$19.5 \\
		Precision (\%) & 93.7$\pm$8.40 & 96.1$\pm$6.32 & 97.4$\pm$5.72 \\
		F1-score (\%) & 67.6$\pm$16.6 & 80.3$\pm$14.5 & 77.8$\pm$14.6 \\
		Averaged AUC & 0.985 & 0.976 & 0.988 \\
		Number of candidates & 4,325 & 2,436 & 9,164 \\
		\hline
	\end{tabular}
\end{table}

Besides ANN, there are several other machine learning methods such as the support vector machine and random forest, which exhibit excellent performance in the pattern recognition. It is worth applying other methods to the pulsar candidate selection and comparing the results. This is beyond the scope of the current paper and will be pursued elsewhere in future.

In this work, we used objects in the Gasperin catalog cross-matched with the ATNF pulsar catalogue and the MILLIQUAS catalog as our training data. Although these are currently the largest available catalogs, the number of cross-matched pulsars is relatively small and we need a further analysis with future larger catalogs to consider if ANN is effective for the selection of pulsar candidates.

Other observable quantities such as the rotation measure and polarization fraction could be useful as inputs. We did not adopted them because the number of radio sources with them are currently very limited. If we can have a sufficient number of samples with polarization data as it is expected in future large surveys, they will make ANNs more effective and narrow the pulsar candidates down further.

\section*{Acknowledgements}
We thank Shiro Ikeda, Shuhei Mano, Shinto Eguchi and Hayato Shimabukuro for useful discussions. NY and SY are financially supported by the Grant-in-Aid from the Overseas Challenge Program for Young Researchers of JSPS. KT is partially supported by Grand-in-Aid from the Ministry of Education, Culture, Sports, and Science and Technology (MEXT) of Japan, No. 15H05896, 16H05999 and 17H01110, and Bilateral Joint Research Projects of JSPS. SY is supported by JSPS KAKENHI Grant Numbers JP16J01585.

\bsp	
\label{lastpage}
\end{document}